# Transverse and Longitudinal Doppler Effects of the Sunbeam Spectra and Earth-Self Rotation and Orbital Velocities, the Mass of the Sun and Others

Sang Boo Nam
7735 Peters Pike, Dayton, OH 45414-1713 USA
sangboonam@mailaps.org

The transverse and longitudinal Doppler effects of the sunbeam spectra are shown to result in the earth parameters such as the earth-self rotation and revolution velocities, the earth orbit semi-major axis, the earth orbital angular momentum, the earth axial tilt, the earth orbit eccentricity, the local latitude and the mass of the sun. The sunbeam global positioning scheme is realized, including the earth orbital position.

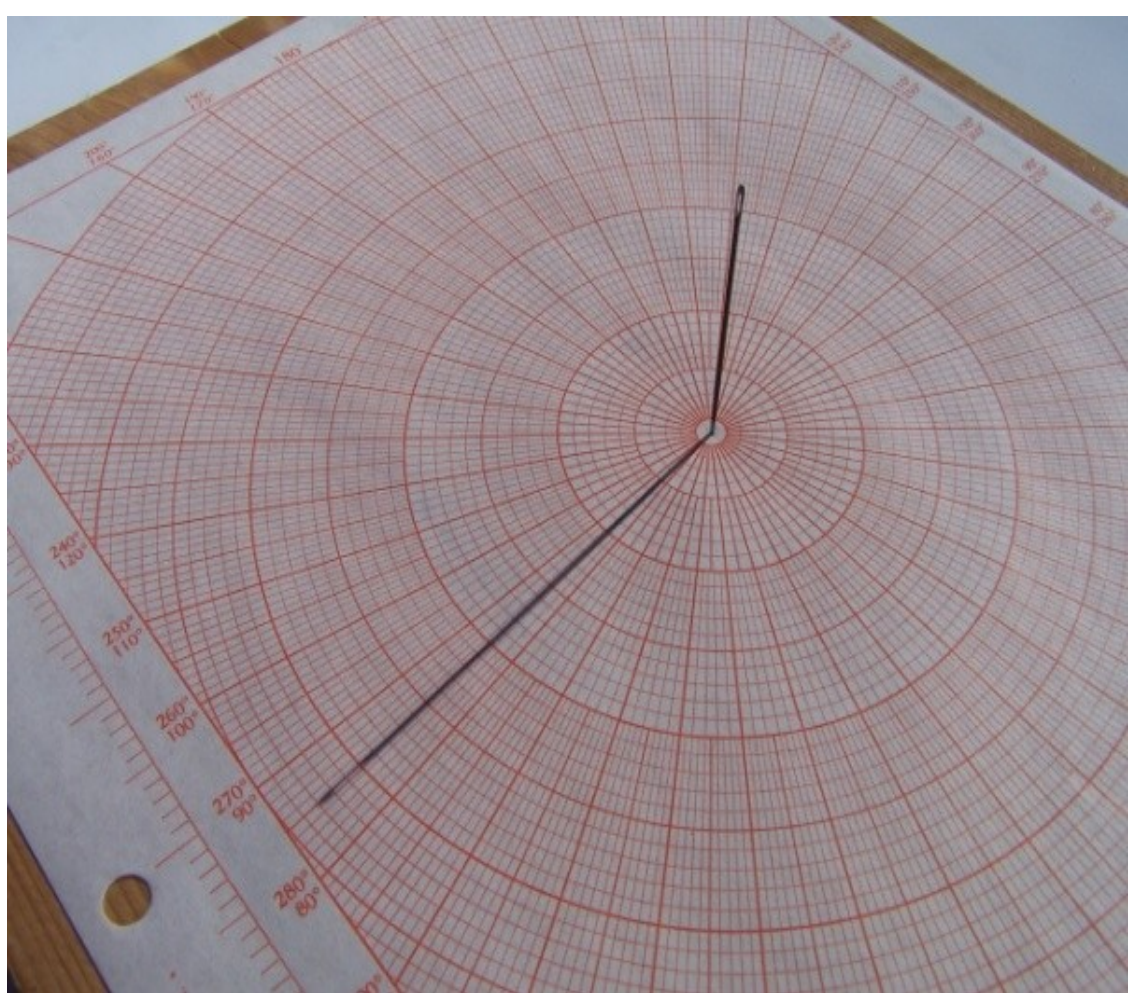

Fig. 1 Sunbeam Global Positioning Device with a needle at the center of radial angle graph paper.

Recently, I showed, using the sunbeam global positioning device shown in Fig. 1, the earth-self rotation and the earth orbiting around the sun and the novel macroscopic wave geometric effect of the sunbeam [1]. I obtained the earth axial tilt Q, the earth orbit eccentricity e and the earth orbital angle $R_S$ at winter solstice and the local latitude u, and made a calendar as well.

However, the earth parameters such as the earth orbit semi-major axis a, the earth-self rotational velocity $V_E$ at the equator and the mass of the sun $M_S$ are assumed to be known. How about earth bound data of them? Yes, I show them.

Here, I present a scheme to obtain the earth orbit revolution velocity, and the earth orbit angular momentum, a, $V_{E,}$ $R_S$, u, Q, e, and $M_S$.

The idea is a simple one, the Doppler shifts. The longitudinal (transverse) Doppler shift goes as linear (square) of the relative velocity between the observer and the signal source. Let me discuss the longitudinal Doppler shift first. In the morning (afternoon), the observer on the earth surface moves toward to (away from) the sun, resulting in the blue (red) shifts of the sunbeam spectra. At noon, the earth-self rotation does not yield any shift, since the earth rotation velocity is perpendicular to the sunbeam direction. The earth orbit radial directional velocity is finite, resulting in the longitudinal Doppler shifts. The transverse Doppler shifts are expected all the time. How large are the Doppler shifts? Numerical values are noted: one electron volt = 2.4 $x10^{14}$Hz. The earth-self rotation velocity at the equator is 0.465 Km/sec = 1.55 $x10^{-6}$ speed of light (C = 3 x $10^5$Km/sec). The average earth orbital velocity is 30 Km/sec = $10^{-4}$C and the earth orbit radial directional velocity of the order of 1.67 x $10^{-6}$C. The transverse Doppler shift of 1 electron volt light (1.24 micron) goes as $10^{-8}$ x electron volt = 2.4 $x10^6$ Hz. The Doppler shifts of the sunbeam spectra are large enough to be observable by the present state of art of technology.

To study the Doppler shift [2], I use the coordinate systems same as those used for the study of the sunbeam directional angle c velocity and recapitulate the results [1]. The bar on the earth surface is considered as the observer. I choose the earth center as the origin of coordinates as shown in Fig. 2. Let the earth-self rotating axis be the z-axis and the earth equatorial plane be the x-y plane. Let the z-y plane be the meridian, noon longitude, and b longitude measured from the meridian and u the latitude. Let the sunbeam from the sun with propagation unit vector come to the earth

$$\mathbf{k = (0, -\cos q, \sin q)} \qquad \mathbf{(1k)}$$

making angle q with the y-axis.

We consider here the spherical earth. The sunset or sunrise longitude g can be determined by the crossing longitude with the sunbeam front plane (sfp), when sfp hits the earth center making angle q with the z-axis. We consider the latitude u circle of radius cosu at z = sinu in the unit of the earth radius. The distance from the z-axis to sfp on the z-y plane is sinu tanq equal to cosu cosg, the projection of the radius cosu on the z-y plane as shown in Fig. 2. Thus, we obtain

$$\cos g = \tan q \tan u. \qquad (1g)$$

This condition is for the radial unit vector

$$\mathbf{r_g = (\cos u \sin g, \cos u \cos g, \sin u)}$$

to be perpendicular to **k**, $\mathbf{k} \bullet \mathbf{r_g} = \mathbf{0}$.

For a given u, the lengths of day and night are determined by the angle q by Eq. (1g). The g = 0 and g = $\pi$ indicate no day and no night, respectively. By Eq. (1g), the north pole region $|u| \geq \pi/2 - |q|$ has no day (no night) for positive (negative) q, and the south pole region the other way around. At equinoxes, q = 0, g = $\pi/2$, we get equal day and night lengths. Strictly speaking, there is a small difference since q is changing from morning to sunset. At winter solstice, q is the earth axial tilt Q. The latitudes $|u| = \pi/2 - |Q|$ are called Arctic and Antarctic circles, respectively. The latitudes u = |Q| are known as Tropic of Cancer and Capricorn, respectively.

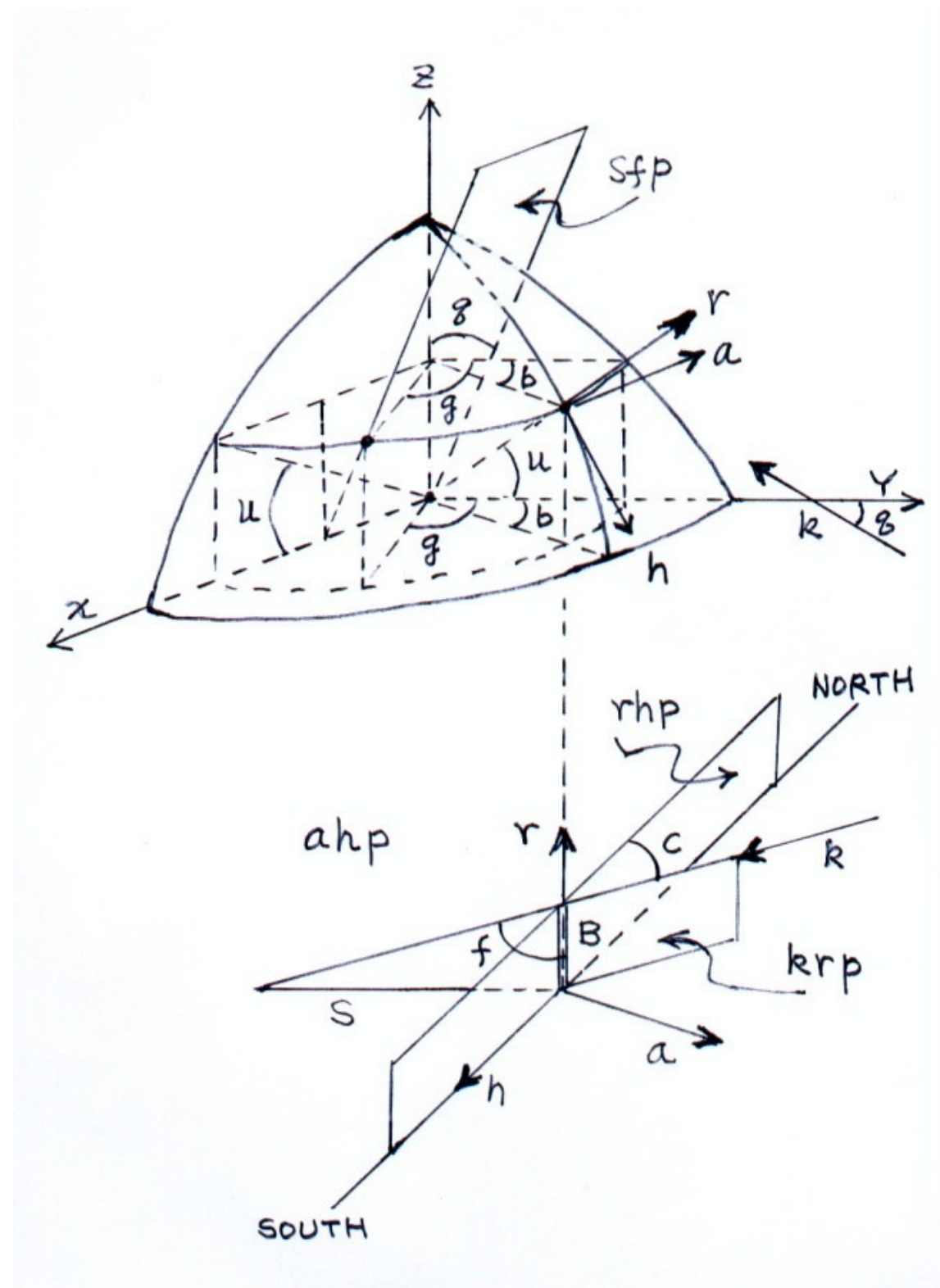

Fig. 2 sfp: sunbeam front plane, krp: **k-r** plane with the sunbeam and a bar, rhp: **r-h** longitudinal plane, ahp: **a-h** plane, angle f between the sunbeam and the bar, angle c between krp and rhp, u: latitude, longitudes b at the bar and g at sunset or sunrise measured from the z-y plane, the meridian, S: length of shadow on ahp, B: length of the bar stands vertically on ahp.

At the place where a bar stands, the radial unit vector **r**, the azimuthal unit vector **a** and the longitudinal unit vector **h** are given by

$$\mathbf{r = (\cos u \sin b, \cos u \cos b, \sin u)}, \qquad (1r)$$

$$\mathbf{a = (-\cos b, \sin b, 0)}, \text{ and} \qquad (1a)$$

$$\mathbf{h = (\sin u \sin b, \sin u \cos b, -\cos u)}. \qquad (1h)$$

We then get the shadow length S by a bar height B on **a-h** plane (ahp), with the angle f between the sunbeam and the bar, as

$$S = B \tan f , \qquad (2)$$

$\cos f = -\mathbf{k} \bullet \mathbf{r} = \cos u \cos q \cos b - \sin u \sin q.$ (3)

The shadow angle c between **k-r** plane (krp) and **r-h** longitudinal plane (rhp), is given by

$\cos c = \mathbf{a} \bullet (\mathbf{k} \times \mathbf{r}) / \sin f$ (4)

$= [\sin u \cos q \cos b + \cos u \sin q]/\sin f$, or

$\tan c = \sin b/[\sin u \cos b + \cos u \tan q].$ (5)

For b = {0; g} = {Noon; Sunrise or Sunset}, we obtain the angles f and c as

$f = \{u + q; \pi/2\},$ (6)

$\tan f = \{\tan(u + q); \text{Infinite}\},$ (7)

$c = \{0; \arccos[\sin q \sec u]\}.$ (8)

By Eqs. (2), (6) and (7), we can examine the earth spherical nature by measuring the shadow lengths at two locations on the same longitudinal line. We expect two different shadow lengths would show the spherical earth. We measure the earth radius $r_E$ with the arc-distance between two locations and the difference between latitudes. At sunrise or sunset, from Eq. (8), the angle c is 90 degrees only at equinoxes, q = 0. Sunrise and sunset directions are not always exactly east and west.

Taking derivatives of angles f of Eq. (3) and c of Eq. (5) with b, we obtain for b = {0; g} = {Noon; Sunrise of Sunset},

$df/dt = BT\partial f/\partial b = \{0 ; \cos u \cos q \sin g\} BT,$ (9)

$dc/dt = BT\partial c/\partial b = \{\cos q \csc(u+q); \sin u\} BT.$ (10)

Here t is the sidereal time and $BT = \partial b/\partial t = 15$ deg/hr, the earth-self rotation angular velocity. There is no term of $\partial q/\partial b$ in Eq. (10) at noon. In other cases, it is negligibly small and neglected in Eq. (9) and Eq. (10). We estimate its size as following. The q changes 4x23.5 degrees in a year, $\partial q/\partial t = 94/365.25$ deg/day = 0.01072 deg/hr and then $\partial q/\partial b = (\partial q/\partial t)/(\partial b/\partial t) < 0.000715$. Its omission is justified in the time frame of a daytime.

The angle c velocity at sunset or sunrise (AcVS) by Eq. (10) is given as

**15 sin(local latitude) deg/sidereal hour**. (11S)

This is same as the rotating rate of swing plane of Foucault pendulum [3], showing the earth-self rotation. The angle c velocity at noon (AcVN) by Eq. (10) is rewritten as

$\mathbf{15\,[\sin u + \cos u \cot(u+g)]\,deg/hr}.$ (11N)

AcVN has an additional term resulted by the novel macroscopic geometric effect of the sunbeam making AcVN faster than AcVS [1].

At the local noon, the shadow has the shortest length and AcVN the maximum. The time difference between the local and standard time noons yields the local longitude = local standard longitude + BT x time difference hours. We have the sunbeam global positioning device.

The earth orbital system around the sun is shown in Fig. 3. I realize the sunbeam direction in the earth system same as the radial direction in the earth orbital system.

$\mathbf{k} = (0, -\cos q, \sin q)$ **is** $\mathbf{r_S} = (\cos R, \sin R, 0)$ (12)

in terms of the respective coordinate variables where the earth orbit angle R is measured from perihelion on the x-axis in the earth orbital system around S-sun. Moreover, the unit vector **w** of the earth-self rotation angular velocity BT in the earth system and its unit vector $\mathbf{w_S}$ in the earth orbital system are given by,

$\mathbf{w} = (0, 0, 1)$ **and** (13)

$\mathbf{w_S} = (\sin Q \cos R_S, \sin Q \sin R_S, \cos Q),$ (14)

where $R_S$ is the earth orbit angle R at winter solstice. Thus, we get easily [4]

$\sin q = \mathbf{k} \bullet \mathbf{w} = \mathbf{r_S} \bullet \mathbf{w_S} = \sin Q \cos(R - R_S)$ or (15q)

$$R = R_S + \arc\cos[\sin q/\sin Q]. \quad (15R)$$

The $R_S$ is obtained by the data of dq/dt at equinoxes by Eq. (19).

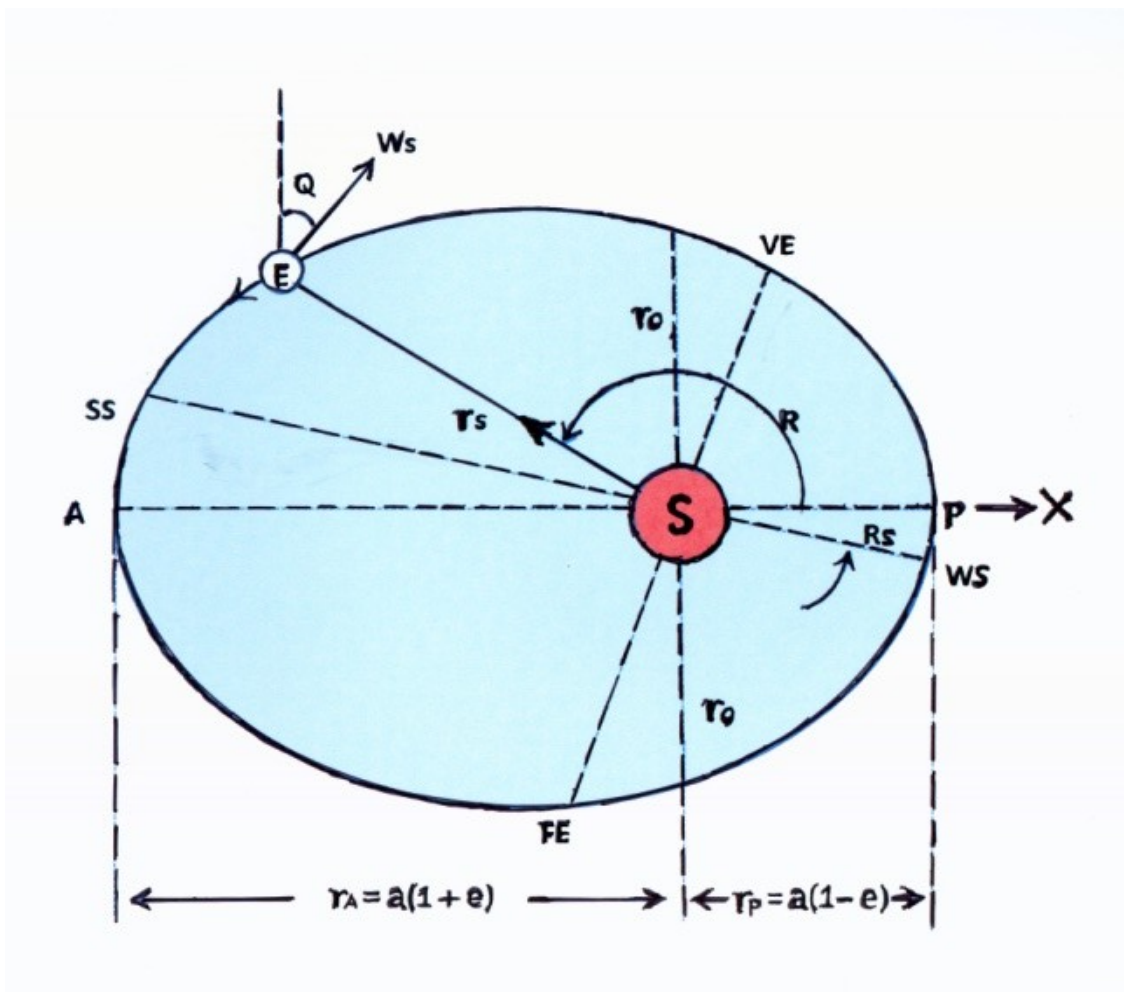

Fig. 3 Earth orbit radial distance $r(R) = r_0/(1 + e \cos R)$, R: the earth orbit angle, $r_0 = a(1 - e^2)$, a and e are the semi-major axis and eccentricity of the ellipse, P: perihelion, VE: vernal (spring) equinox SS: summer solstice, A: aphelion, FE: fall equinox, WS: winter solstice, $R_S$: angle at WS, Q: the earth axial tilt, $\mathbf{W_S = (\sin Q \cos R_S, \sin Q \sin R_S, \cos Q)}$: BT unit vector, $\mathbf{r_S = (\cos R, \sin R, 0)}$: the r(R) unit vector.

By Eq. (15q), I get the angle R velocity as

$$dR/dt = -(dq/dt) \cos q \csc Q \csc(R - R_S). \quad (16q)$$

This equals to that by the earth orbit [3]

$$dR/dt = L/M_E r^2(R)$$

$$= (L/M_E r_0^2)(1 + e \cos R)^2, \quad (16R)$$

$$r(R) = r_0/(1 + e \cos R), \quad (16r)$$

$$r_0 = a(1 - e^2) = L^2/M_E^2 GM_S, \quad (16a)$$

$$L/M_E r_0^2 = 2\pi/Y(1 - e^2)^{3/2}, \text{ and} \quad (16L)$$

$$Y = 2\pi a^{3/2}/(GM_S)^{1/2}. \quad (16Y)$$

Here Y is the Kepler period of the elliptic orbit [5], L the earth orbital angular momentum, $M_E$ the earth mass, G the Newton gravitational constant, and r(R) the earth orbit radial distance from the sun, respectively. We consider the function Z(R) defined, for $|R| \le \pi/2$, by

$$Z(R) = [(dR/dt)(R)/(dR/dt)(R+\pi)]^{1/2}$$

$$= r(R + \pi)/r(R) \quad (17r)$$

$$= (1 + e \cos R)/(1 - e \cos R) \quad (17R)$$

$$= |(dq/dt)(R)/(dq/dt)(R + \pi)|^{1/2}. \quad (17q)$$

Z(R) has a maximum value at R = 0, which can be found. By Eq. (17R) and Eq. (17q), we get

$$e = [Z(0) - 1]/[Z(0) + 1]. \quad (18)$$

The eccentricity of the earth orbit is obtained with data of dq/dt at perihelion and aphelion. Moreover, by Eq. (15q) and Eq. (16q), dq/dt has local extreme values at equinoxes (q = 0), which can be found. For $Z(R_S + \pi/2) = Z_E$, we obtain

$$R_S = \arc\sin[(1 - Z_E)/e(Z_E + 1)]. \quad (19)$$

The angle $R_S$ at winter solstice is obtained with data of dq/dt at equinoxes. By integrating dR/dt of Eq. (16R), we obtain a calendar, D days from perihelion as

$$D = (Y/2\pi)[H - eE^{1/2} \sin R/(1 + e \cos R)], \quad (20)$$

$$H = 2 \arc\tan[K \tan(R/2)],$$

$$E = 1 - e^2, \text{ and}$$

$$K = [(1 - e)/(1 + e)]^{1/2}.$$

R and q are related by Eq. (15q) and Eq. (15R). The second term in Eq. (20) shows the earth elliptic orbit.

The Doppler shift formula for frequencies $W_1$ and $W_2$ in the systems 1 and 2 is given by [6]

$$W_2 = \text{gamma} W_1 [1 - \text{beta}] \quad (21)$$

$$\text{beta} = \mathbf{k}\bullet\mathbf{V}_{21}/C \qquad (21b)$$

$$\text{gamma} = [1 - (\mathbf{V}_{21}/C)^2]^{-1/2}, \qquad (21g)$$

$$\mathbf{V}_{21} = (dr/dt)\mathbf{r}_S + (rdR/dt)\mathbf{v}_S + (V_E \cos u)\mathbf{a} \qquad (21v)$$

$$\mathbf{v}_S = \mathbf{e}_Z \times \mathbf{r}_S, \ \mathbf{r}_S \times \mathbf{v}_S = \mathbf{e}_Z = (0, 0, 1)_S,$$

where $\mathbf{V}_{21}$ is the relative velocity between the systems 1 and 2. The unit vector $\mathbf{e}_Z$ is the z-direction in the earth orbital system, and its components ($e_1$, $e_2$, $e_3$) in the earth system are determined by the following conditions

$$\mathbf{k}\bullet\mathbf{e}_Z = 0 = \mathbf{r}_S\bullet\mathbf{e}_Z, \ \mathbf{w}\bullet\mathbf{e}_Z = \mathbf{e}_3 = \mathbf{e}_Z\bullet\mathbf{w}_S = \cos Q. \qquad (21z)$$

The beta (gamma) results in the longitudinal (transverse) Doppler shift.

Using Eq. (16R), Eq. (16r), Eq. (1k) and Eq. (1a), beta is obtained as

$$\text{beta} = [V e \sin R - V_E \cos u \sin b \cos q]/C, \qquad (22)$$

$$V = L/M_E r_0, \qquad (23)$$

$$V_E = r_E \, BT. \qquad (24)$$

The first term in Eq. (22) is the earth orbit radial directional velocity and the second the earth-self rotation velocity projected on **k**, **k•a** $V_E \cos u$. At the local noon, b = 0, beta has only the first term. From perihelion to vernal equinox and to aphelion, we have the Doppler red shifts and the Doppler blue shifts in the rest region.

First we consider the longitudinal Doppler shift and the sunbeam frequency F(P) at perihelion at noon (R=0, beta=0) as the reference frequency, disregarding the gamma factor. We write the frequency shift at R as

$$W(R) = F(R) - F(P) = F(P)\text{beta}, \qquad (25W)$$

W(VE) for at vernal equinox $R=R_{VE}=R_S+\pi/2$, etc.. Then, by Eq. (22) we get at noon, b = 0,

$$Ve = C\,[W(\pi/2) - W(3\pi/2)]/2, \qquad (25V)$$

$$\tan R_S = W(WS)/W(VE) = W(SS)/W(FE). \qquad (25R_S)$$

Using Eq. (16a), Eq. (16L) and Eq. (23), we get

$$a = V\,(1 - e^2)^{1/2} Y/2\pi, \qquad (27a)$$

$$L = V^2\,(1 - e^2)^{3/2} M_E\,Y/2\pi, \text{ and} \qquad (27L)$$

$$M_S = V^3\,(1 - e^2)^{3/2} Y/2\pi G. \qquad (27M_S)$$

Using Eq. (1g), the second term in Eq. (22) may be rewritten at sunrise (SR) or sunset (SS) as

$$- V_E\,[\cos(u+q)\cos(u-q)]^{1/2}/C.$$

Neglecting small change in the first term in Eq. (22) during a daytime period, we get

$$V_E = C|W(SR) - W(SS)|(u=0,q=0)/2. \qquad (27V_E)$$

The earth-self rotation angular velocity BT is obtained by $BT = V_E/r_E$, with the earth radius $r_E$. Using Eq. (25V) and Eq. ($27V_E$), we get

$$[\cos(u+q)\cos(u-q)]^{1/2} \text{ (any day)}$$

$$= C|W(SR)-W(SS)|/2V_E \qquad (28)$$

$$\sin R = C[W(R \text{ at Noon}) - W(P \text{ at Noon})]/Ve. \qquad (29)$$

By Eq. (15q), Eq. (28) and Eq. (29), u and q are obtained. At winter solstice q = Q is obtained.

We now discuss the transverse Doppler shift. For gamma of Eq. (21g), the relative velocity square is needed. For simplicity, we present the square of relative velocity at noon at perihelion as

$$(V_{21})^2 = V^2\,(1+e)^2 + (V_E \cos u)^2$$

$$-2VV_E\,(1+e)\cos u \cos Q \sec q_P, \qquad (30)$$

$$\sin q_P = \sin Q \cos R_S \quad \text{from Eq. (15q).}$$

The first term comes from the earth orbital revolution velocity and the second term from the earth rotational velocity. The last term comes from the scalar product of two velocities. If the sunbeam spectra in the rest frame are known by measuring He-lines or H-lines, we get the transverse Doppler shift at perihelion as

$$W_T(P) = |F(P) - F(rest)| = F(P)(V_{21}/C)^2/2 . \quad (31)$$

By Eq. (25V), Eq. (30) and Eq. (31), the earth orbit revolution velocity V and e are obtained. At aphelion, we get $W_T(A)$ by Eq. (30) with -e and $q_A = -q_P$. We get the frequency difference

$$|F(A) - F(P)| = F(P)|(V_{21})_P{}^2 - (V_{21})_A{}^2|/2C^2$$

$$= F(P)2e(V/C)^2[1-(V_E/V)\cos u \cos Q \sec q_P]. \quad (32)$$

Here we used $\sec q_A = \sec q_P$ by Eq. (15q). By Eq. (25V) and Eq. (32), V and e are obtained.

I stress Eq. (15q) and Eq. (21z) play crucial roles for making connections between variables in the earth and orbital systems, and obtaining a calendar by q measured on the earth surface and the Doppler shifts of the sunbeam spectra.

The Doppler shifts of the sunbeam spectra result in the parameters $R_S$, V, $V_E$ and e without any parameter. One input variable Y is needed to get the earth orbit semi-major axis and the sun mass with G. For the earth orbit angular momentum, $M_E$ and Y are needed. Nowadays, $M_E$ is obtained by the satellite orbit semi-major axis and its period revolving around the earth.

Moreover, the global positioning scheme is realized, including the earth orbit position on the local date, without any satellite.

It is highly desirable to measure the Doppler shifts of the sunbeam spectra. Specially, the transverse Doppler shift of the sunbeam spectra would be physically one of challenging works.

**Photon has so many beautiful sides** [7]**.**

References

[1] S. B. Nam, arXiv: 0910.5767 (physics.gen-ph).

[2] Historical notes are in order for the transverse Doppler shift. To prove the Einstein special theory of relativity, Ives and G. R. Stilwell, J. Opt. Soc. Am, **28**, 215 (1938); ibid, **31**, 369 (1941) observed radiations from the atomic beam and other directions, and proved the time dilation. Using the Mossbauer effect, [Z. Physik A, 151, 124 (1958)], Hay, et al, Phys. Rev. Lett., **4**, 165 (1960) and Kundig, Phys. Rev. 129, 2371 (1963) used gamma ray to show the transverse Doppler effect. Greenberg. et al, Phys. Rev. Lett. **23**. 1267 (1969); (E), ibid, **23**, 1473 (1969) observed the time dilation by charged pion life times at the rest and in flight. Flying with the atomic clock eastward and westward trips, Hafele and Keating, Science, 177, 166 (1972) measured directly the time dilation. Nowadays, the global positioning system and weather broad castings are based on the Doppler shifts with satellites, and many others. Of course, the astronomy data are based on the Doppler shifts.

[3] L. D. Landau and E. M. Lifshitz, Mechanics, (Pergamon Press, New York 1960).

[4] In general BT has the factor for precession of equinoxes with the period of 26000 years, which is not interested here. Writing $\mathbf{w_S = (A, B, C)}$, we get $\sin q = \mathbf{k \bullet w = r_S \bullet w_S = A \cos R + B \sin R}$. Conditions of q = Q at winter solstice and q = 0 at vernal equinox, yield A and B, and then Eq. (15q) and Eq. (15R).

[5] We obtain the Kepler period of the elliptic orbit $Y = 2\pi a^{3/2}/(GM_s)^{1/2} = 365.2440$ days with the major semi-axis $a = r_0/(1 - e^2) = 1.49607E11$m and $GM_s = 1.327461E20\ m^3/s^2$. The Gregorian calendar year is 365.2425 solar days. The leap year is every four years, except every hundred year, but every 400 years.

[6] J. D. Jackson, Classical Electrodynamics, (John Wiley & Sons, 2nd Ed. New York, 1975).

[7] A. Aiello, N. Lindlein, C. Marquardt, and G. Leuchs, Phys. Rev. Lett. 103, 100401 (2009) presented Transverse Angular Momentum and Geometric Spin Hall Effect of Light occurring when a polarized beam of light is observed from a reference frame tilted with respect to the direction of propagation of the beam.